\documentclass[preprintnumbers,amsmath,amssymb,floatfix,11pt,onecolumn,prd,superscriptaddress,nofootinbib,showpacs,showkeys]{revtex4}
\usepackage{graphicx}
\usepackage{epsfig}
\usepackage{bm}
\usepackage{amsfonts}

\def\ben{\begin{equation}}
\def\een{\end{equation}}

 \def\bd{\begin{document}} \def\ed{\end{document}}
\def\ds{\documentstyle} \let\fr=\frac \let\bl=\bigl \let\br=\bigr
\let\Br=\Bigr \let\Bl=\Bigl
\let\bm=\bibitem
\let\na=\nabla
\let\pa=\partial \let\ov=\overline
\newcommand{\be}{\begin{equation}}
\newcommand{\ee}{\end{equation}}
\def\ba{\begin{array}}
\def\ea{\end{array}}
\def\ft#1#2{{\textstyle{\frac{\scriptstyle #1}{\scriptstyle #2} } }}
\def\fft#1#2{{\frac{#1}{#2}}}
\def\del{\partial}
\def\vp{\varphi}
\def\sst#1{{\scriptscriptstyle #1}}
\def\oneone{\rlap 1\mkern4mu{\rm l}}
\def\td{\tilde}
\def\wtd{\widetilde}
\def\ie{{\it i.e.\ }}
\def\dalemb#1#2{{\vbox{\hrule height .#2pt
        \hbox{\vrule width.#2pt height#1pt \kern#1pt
                \vrule width.#2pt}
        \hrule height.#2pt}}}
\def\square{\mathord{\dalemb{6.8}{7}\hbox{\hskip1pt}}}
\newcommand{\ho}[1]{$\, ^{#1}$}
\newcommand{\hoch}[1]{$\, ^{#1}$}
\newcommand{\bea}{\setlength\arraycolsep{2pt} \begin{eqnarray}}
\newcommand{\eea}{\end{eqnarray}}
\newcommand{\ra}{\rightarrow}
\newcommand{\lra}{\longrightarrow}
\newcommand{\Lra}{\Leftrightarrow}
\newcommand{\bp}{\tilde \beta^\prime}
\newcommand{\tr}{{\rm tr} }
\newcommand{\Tr}{{\rm Tr} }
\def\0{{\sst{(0)}}}
\def\1{{\sst{(1)}}}
\def\2{{\sst{(2)}}}
\def\3{{\sst{(3)}}}
\def\4{{\sst{(4)}}}
\def\5{{\sst{(5)}}}
\def\6{{\sst{(6)}}}
\def\7{{\sst{(7)}}}
\def\8{{\sst{(8)}}}
\def\m{{\sst{(m)}}}
\def\n{{\sst{(n)}}}
\def\cA{{{\cal A}}}
\def\cB{{{\cal B}}}
\def\cF{{{\cal F}}}
\def\cG{{{\cal G}}}
\def\cH{{{\cal H}}}
\def\tV{\widetilde V}
\def\tW{\widetilde W}
\def\tH{\widetilde H}
\def\tE{\widetilde E}
\def\tF{\widetilde F}
\def\tA{\widetilde A}
\def\im{{{\rm i}}}
\def\tY{{{\wtd Y}}}
\def\ep{{\epsilon}}
\def\vep{{\varepsilon}}
\def\bD{{{\bar D}}}
\def\R{{{\mathbb R}}}
\def\C{{{\mathbb C}}}
\def\H{{{\mathbb H}}}
\def\CP{{{\mathbb C}{\mathbb P}}}
\def\RP{{{\mathbb R}{\mathbb P}}}
\def\Z{{{\mathbb Z}}}
\def\bA{{{\mathbb A}}}
\def\bB{{{\mathbb B}}}
\def\bC{{{\mathbb C}}}
\def\bD{{{\mathbb D}}}
\def\bE{{{\mathbb E}}}
\def\bZ{{{\mathbb Z}}}
\def\Re{{{\frak{Re}}}}
\def\Im{{{\frak{Im}}}}
\def\cosec{{\,\hbox{cosec}\,}}
\def\Gm{{\Gamma_{\!\! -}}}
\def\Gp{{\Gamma_{\!\! +}}}
\def\stan{{standard }}
\def\nonstan{{supernumerary }}
\def\p{{\partial}}
\def\kdel#1{{\fft{\del}{\del#1}}}

\def\bog{{Bogomolny }}
\def\om{{\omega}}

\newcommand{\nnr}{\nonumber \\}
\newcommand{\pd}{\partial}
\newcommand{\ud}{\textrm{d}}
\newcommand{\dTH}{T^{\prime \, 0}_\textrm{H}}
\newcommand{\dOi}{\Omega^{\prime \, 0}_i}
\newcommand{\bx}{{\bf x}}
\begin{document}

\title{Condensation of the scalar field with Stuckelberg and Weyl Corrections in the background of a planar AdS-Schwarzschild black hole}
\author{\textbf{ D. Momeni}}
 \affiliation{Eurasian International Center
for Theoretical Physics, Eurasian National University, Astana
010008, Kazakhstan}
\author{\textbf{ M. R. Setare}}
\affiliation{Department of Science, Payame Noor University, Bijar, Iran}
\author{\textbf{ Ratbay Myrzakulov}}
\affiliation{Eurasian International Center
for Theoretical Physics, Eurasian National University, Astana
010008, Kazakhstan}

\begin{abstract}
We study analytical properties of the Stuckelberg holographic  superconductors
with Weyl corrections. We obtain the minimum critical temperature as a function of the mass of the
 scalar field $m^2$. We show that in limit of the $m^2=-3$,$T^{Min}_c\approx0.158047\sqrt[3]{\rho }$
  which is close to the numerical estimate $T_c^{Numerical}\approx 0.170\sqrt[3]{\rho}$. Further we
   show that the mass of the scalar field in bounded from below by the
$  m^2>m_c^2$ where $m_c^2=-5.40417$. This lower bound is weaker and different from the previous
 lower bound $m^2=-3$ predicted by stability analysis. We show that in the Breitenlohner-Freedman bound,
  the critical temperature remains finite. Explicitly, we prove that here there is exist
  a linear relation between $<O_{\Delta}>$ and the chemical potential.

\end{abstract}
\pacs{04.70.Bw, 11.25.Tq, 74.20.-z}
 \keywords{Classical Black holes; Gauge/string duality; High-$T_C$ superconductors theory}
 \newpage
 \maketitle
\section{Introduction}
The anti de Sitter/conformal field theory (AdS/CFT) correspondence
conjecture \cite{maldacena} is a very powerful tool for condensed matter physics specially for critical behavior of systems(see for instance
\cite{condencesd1,condencesd2} and references therein) in high temperature superconductors \cite{super1,super2}.
Different kinds of the holographic superconductors have been studied
in Einstein theory \cite{GR1,GR2} or extended versions such as
Gauss-Bonnet (GB)\cite{GB1,GB2,GB3,GB4,GB5} and even in
Horava-Lifshitz theory \cite{HL1,HL2}. Further the effect of
 magnetic fields on superconductors have been discussed \cite{wen1,wen2,epl1}.
  There are some other types of superconductors with non linear Maxwell fields \cite{born1,born2}
or with Chern-Simon terms \cite{cs}. Also holographic superconductor models with the Maxwell field
 strength corrections have been investigated \cite{maxwell}. Even, recently the holographic approach
  has been used for Josephson Junction effect \cite{cai2012}.\\
 AdS/CFT can also describe
superfluid states in which the condensing operator is a vector and
hence rotational symmetry is broken, such states are termed  p-wave
superfluid states \cite{pwave1,pwave2,pwave3}. Here the CFT has a
global $SU(2)$ symmetry and hence three conserved currents
$J^{\mu}_a$ , where $a = 1, 2, 3$ label the generators of $SU(2)$.
 Many of these works are based on a numerical analysis of the equations of motion
(EOM) near the horizon and the asymptotic limit by a suitable
shooting method. The pioneering work on analytic methods in this
topic was by Hertzog \cite{herzog}. He showed that at least in probe
limit, by solving equations analytically (the  perturbation theory),
one can obtain the critical exponent and the expectation values of
the dual operators.
 Near the critical point the value of the
scalar field $\psi$ is small and consequently we can treat the
expectation values of the dual boundary operators
$\epsilon\equiv<O_{\Delta_{\pm}}>$ as a perturbation parameter. This
method has been used recently by Kanno for investigating the GB
superconductors even away from the probe limit \cite{kanno}.
Applying the analytical methods has lead to new trends (see for
example \cite{analytic1,analytic2} and the references in it). There is a much more beatiful variational method to study the critical behavior of holographic superconductors
\cite{analytic1,analytic2}.  Instead of
using shooting numerical algorithms, we can obtain the critical
temperature $T_c$ and the exponent of the criticality
 by computing a simple variational approach. They studied different
  modes of super criticality
s-wave, p-wave and even d-wave. Thus as we know, there are two major
methods for analytical study of superconductors:

\emph{1- The
small parameter perturbation theory}  \cite{herzog}
 \emph{2- The Sturm�Liouville  variational method} \cite{analytic1,analytic2}.
  We must mention here
that, the variational method, which has been used in the present
work, gives only the minimum value of the critical temperature
$T^{Min}_c$ for a model with a typical parameter. For example if
we focus on Weyl corrections to holographic superconductors, as
 shown in \cite{weyl}, for a large range of the
coupling value $-\frac{1}{16}<\gamma<\frac{1}{24}$, there is a
universal relation for the critical temperature $T_c\simeq
\sqrt[3]{\rho}$. The proportionality constant depends on the Weyl
coupling $\gamma$ and can be computed . In this case we found that
temperature $T^{Min}_c=0.170\sqrt[3]{\rho}$ corresponds to the
value $\gamma=-0.06$. In a recent paper, we showed that this critical
 temperature can be obtained from the variational method \cite{mpla}.\\
Recently there is much interest on GB and Weyl corrected and
specially the Stuckelberg superconductors, even in the presence of
the external magnetic fields \cite{jian2010}. Recently, we
investigated the p-wave holographic superconductors with Weyl
corrections \cite{epl2}. The Stuckelberg holographic superconductors
with Weyl corrections have been studied recently \cite{plb2011}.
They studied the problem numerically. Our program in this paper is
studying the Weyl corrections to the Stuckelberg superconductors
analytically.

Our plan is organized as follows. In section 2, we construct the
basic model of the $3+1$ holographic superconductor with Weyl
corrections. In section 3 we present the analytical results for the
condensation and minimum value of the critical temperature for
different scaling and  the critical exponent $\beta$ via variational
bound. Conclusions and discussions follow in section 4.

\section{Weyl corrected Stuckelberg holographic superconductors}

The s-wave Stuckelberg  holographic superconductors constructed from
an Abelian $U(1)$ gauge field coupled to a massive charged  (complex) scalar field.
 The simplest form of the action in five dimensions
($3+1$ holographic picture)  with Weyl corrections is \cite{plb2011}

\begin{eqnarray}
 S=\int dtd^{4}x\sqrt{-g}(R+\frac{12}{l^2}-L).\label{action}
\end{eqnarray}
with modified matter Lagrangian and Stuckelberg potental function
$F(\psi)$, \be L=\frac{1}{4}(F^{\mu\nu}F_{\mu\nu}-4\gamma
C^{\mu\nu\rho\sigma}F_{\mu\nu}F_{\rho\sigma})+
\frac{\partial_{\mu}\psi\partial^{\mu}\psi}{2}+\frac{m^2\psi^2}{2}+\frac{1}{2}F(\psi)A_{\mu}A^{\mu}.
\ee The general form of the function $F(\psi)$ is
\begin{eqnarray}\label{F}
F(\psi)=\psi^2+c_{\alpha}\psi^{\alpha}+c_4\psi^4.
\end{eqnarray}
Here $3\leq\alpha\leq4$ and $c_4,c_{\alpha}$ are two constants of
order $O(1)$. We write the action in units  $2\kappa^2=1,$ in which
the anti de-Sitter
 (AdS) radius $ l=1$, the negative cosmological constant in this units
is just $12$, charge $e=1$,
$F_{\mu\nu}=\partial_{\mu}A_{\nu}-\partial_{\nu}A_{\mu}$. The gauge
field $A_{\mu}$ lives in bulk and produces a conserved current
$J^{\mu}$. This current corresponds to a global $U(1)$ symmetry. For
this reason, $\psi$ is a real function.
 About the action (\ref{action}) we can
say that \emph{since the background geometry will be an Einstein
metric, we will argue that there is a unique tensorial structure
correcting the Maxwell term at leading order in derivatives, arising
from a coupling to the Weyl tensor and leading to the dimension-six
operator in (1) parametrized by the constant $\gamma$. Other
curvature couplings simply provide constant shifts when considering
linearized gauge field fluctuations about the background. There is
another reason for considering the Weyl correction, which is related
to the quantum corrections. In any background in which additional
charged matter fields are integrated out below their mass threshold,
the Weyl coupling $C^{\mu\nu\rho\sigma}F_{\mu\nu}F_{\rho\sigma}$ is
generated at 1-loop, with a coefficient $\gamma=\frac{\alpha}{m^2}$}
first computed (for four dimension) by Drummond and Hathrell
\cite{qc}. The Weyl's coupling $\gamma$ is limited since its value
lies in the interval $-\frac{1}{16}<\gamma<\frac{1}{24}$. In the
probe limit, we neglect from the back reactions and in this case,
the gravity sector is effectively decoupled from the matter field's
sector. In this probe limit, the exact solution for Einstein
equations is a planer AdS-Schwarzschild black hole given by
\begin{eqnarray}
 ds^2=r^2(-fdt^2+dx^idx_i)+\frac{dr^2}{r^2f}\label{g},
\end{eqnarray}
here
\begin{eqnarray}
f=1-(\frac{h}{r})^4,
\end{eqnarray}
and the horizon is located at $r=h$. This solution is asymptotically
anti-de Sitter. The temperature of the dual conformal field theory
 is nothing but the Hawking temperature
$T=\frac{h}{\pi}$.
 The numerical analysis of the phase transition for this model has been discussed recently \cite{plb2011}.
 When the temperature of the black hole falls
below a critical value $T_c$, a phase transition occurs between the
normal phase and a new phase, in which the scalar field $\psi$
condenses. If the model has this solution, we conclude that our
field theory has a superfluid phase. We choose a gauge as
$\psi=\psi(r),A_{t}=\varphi(r)$. It is more convenient to work in
terms of the dimensionless parameter $\xi=\frac{h}{r}$, in which the
horizon is $\xi=1$ and the boundary at infinity located at $\xi=0$.
The resulting equations for metric (\ref{g}) are given by
\cite{plb2011}
\begin{eqnarray}
 \psi''+\Big(\frac{f'}{f}+\frac{5}{r}\Big)\psi'+\frac{\phi^2}{2r^4 f^2}\frac{dF}{d\psi}-\frac{m^2\psi}{r^2 f}=0\label{eom1}\\ \ \
 \Big(1-\frac{24\gamma h^4}{r^4}\Big)\phi''+\Big(\frac{3}{r}+\frac{24\gamma h^4}{r^5}\Big)\phi'-\frac{F}{r^2 f}\phi=0\label{eom2}
 \end{eqnarray}
 where prime  denotes derivative with respect to $r$. In terms of the $\xi$, Eqs.
  (\ref{eom1}), (\ref{eom2}) become
 \begin{eqnarray} \frac{d}{d\xi}\Big(\xi^2\frac{d\psi}{d\xi}\Big)-\frac{\xi(5-\xi^4)}{1-\xi^4}\frac{d\psi}{d\xi}+\frac{\phi^2\xi^2}{2h^2(1-\xi^4)^2}\frac{dF(\psi)}{d\psi}-\frac{m^2}{1-\xi^4}\psi=0,
 \label{eom11}\\ \ \
 (1-24\gamma\xi^4)\frac{d}{d\xi}\Big(\xi^2\frac{d\phi}{d\xi}\Big)-3\xi(1+8\gamma\xi^4)\frac{d\phi}{d\xi}-\frac{F(\psi)\phi}{1-\xi^4}=0
\label{eom22}
 \end{eqnarray}
 It is helpful to compare the analytical results with numerical results presented in \cite{plb2011}.
  For stability of the theory, we
 fix the mass of the scalar field
 above the  Breitenlohner-Freedman (BF) bound \cite{BF}. This BF bound is a lower bound on
  the mass of the scalar field $m^2$. For a theory defined in $d+1$ dimensions,
  the upper bound on the mass reads $m_{BF}^2=-\frac{d(d-1)}{4}$.
  This bound comes from the stability analysis. In this paper, from holographic picture
   of s-wave superconductors we will present another upper bound independent from this
    bound. Back to our model, the adequate and sufficient boundary conditions for these equations
can be written on horizon $\xi=1$, the bulk's boundary $\xi=0$. On
the horizon we have $\varphi(1)=0,\psi'(1)=\frac{2}{3}\psi(1)$ and
on the boundary of bulk, the asymptotic forms of the solutions are
\begin{eqnarray}
 \varphi\approx \mu-\frac{\rho}{h} \xi^2\\
 \psi
 \approx\epsilon\xi^{\Delta_{\pm}}
 =\psi^{(1)}\xi^{\Delta_{+}}+\psi^{(3)}\xi^{\Delta_{-}}\label{aproxpsi},
 \end{eqnarray}
 here
  $$\epsilon=\frac{<O_{\Delta_{\pm}}>}{\sqrt{2}h^{\Delta_{\pm}}}\label{epsilon}
  $$
$\mu$ and $\rho$ are dual to the chemical potential and charge
density of the boundary CFT, $\psi^{(1)}$ and $\psi^{(3)}$ are dual
to the source and expectation value of the boundary operator $O$
respectively. $<O_{\Delta_{\pm}}>$ are the condensation with
dimension $\Delta_{\pm}$ where \be \Delta_{\pm}\equiv
\Delta=\{3,1\}. \ee The conformal scaling dimension $\Delta\geq1$ is
related to the mass and the de Sitter radius by
$m^2=\triangle(\triangle-4)$.

\section{Analytical results for the condensation and critical temperature}

We know that there is a second order continuous phase transition at
the critical temperature, the solution of Eq.(\ref{eom22}) at $T_c$
is
 \be
\phi=\lambda h_c(1-\xi^2)
\ee
where $h_c$ is the radius of the horizon at $T= T_c$.
 As $T\rightarrow T_c$, the scalar filed's equation of motion (\ref{eom11}) takes  the
 following form
\begin{eqnarray} -\psi''+g(z)\psi'+\frac{m^2}{\xi^2(1-\xi^4)}\psi=\frac{\lambda^2}
{2(1+\xi^2)^2}\frac{dF}{d\psi},\label{psi}
\end{eqnarray}
here $\lambda=\frac{\rho}{h_c^{3}}$, and
$$g(\xi)=\frac{\xi^4+3}{\xi(1-\xi^4)}$$

 By solving the equation
(\ref{psi}), we can obtain the value of $T_c$. To match the behavior at
the boundary, we can define
\begin{eqnarray}
  \psi(\xi)=
  \epsilon\xi^{\Delta}\Omega(\xi)
 \end{eqnarray}
where, according to Eq.(\ref{eom11}), $\Omega$ is normalized as
$\Omega (0) = 1$. We deduce
\begin{eqnarray}
-\Omega''+(g(\xi)-\frac{2\Delta}{\xi})\Omega'+(\frac{\Delta g(\xi)}{\xi}-\frac{\Delta(\Delta-1)}{\xi^2}+\frac{m^2}{\xi^2(1-\xi^4)})\Omega=\frac{\lambda^2 \xi^{-\Delta}}{2(1+\xi^2)^2}\frac{dF}{d\psi}|_{ \psi(\xi)=
 \epsilon\xi^{\Delta}\Omega(\xi)}\label{omega}
 \end{eqnarray}
 when $\xi\rightarrow 0$,
$\frac{\Omega'}{\xi}$ should be finite, so this equation is to be
solved subject to the auxiliary boundary condition $\Omega'(0)=0$.

\subsection{Variational approach}

 Now we use the variation method to solve the Sturm�Liouville (S-L) problem. The
(S-L) eigenvalue problem is to solve the equation (\ref{omega})
\begin{eqnarray}
\frac{d }{d\xi}[k(\xi) \frac{d\Omega}{d\xi}  ] - q(\xi)\Omega(\xi)
+ \frac{\lambda^2\rho(\xi)}{2}\Phi'(\Omega(\xi)) = 0
 \end{eqnarray}
 with boundary condition
\begin{eqnarray}
k(\xi)\Omega(\xi)\Omega'(\xi)|^1 _0 = 0\label{bc}
\end{eqnarray}
 The (S-L)
problem can be converted to a functional minimize problem
\begin{eqnarray}
F [\Omega(\xi)] = \frac{\int ^1 _0 d\xi(k(\xi)\Omega ' (\xi) ^2 +
q(\xi)\Omega(\xi) ^2)}{\int ^1 _0 d\xi\rho(\xi)\Phi(\Omega(\xi)) }\label{functional}
 \end{eqnarray}
Then  $n'$th eigenvalue $\lambda_n$ can also be obtained by
variation of Eq. (\ref{functional}). This eigenvalue is the minimum
value of a sequence of the eigenvalues $\{\lambda_n\}^{\infty}_{0}$
i.e. we obtain $\lambda_{0}<\lambda_n$. It is a familiar result from
the functional theory. For Eq.(\ref{omega}) we immediately obtain
\begin{eqnarray}
k(\xi)=\xi^{2\Delta-3}(1-\xi^4)
\\
q(\xi)=-k(\xi)(\frac{\Delta g(\xi)}{\xi}-\frac{\Delta(\Delta-1)}{\xi^2}+\frac{m^2}{\xi^2(1-\xi^4)})
\\
\rho(\xi)=\xi^{\Delta-3}\frac{1-\xi^2}{1+\xi^2}\\
\Phi(\Omega(\xi))\equiv F(\Omega(\xi))=
 ( \epsilon\xi^{\Delta}\Omega(\xi))^2\Big(1+c_4  ( \epsilon\xi^{\Delta}\Omega(\xi))^2+c_{\alpha} ( \epsilon\xi^{\Delta}\Omega(\xi))^{\alpha-2}\Big)
 \end{eqnarray}
We will follow this method in next section.

\subsection{Analytical results for $\Delta=3$}
If we fix $\Delta=3$, then the boundary condition (\ref{bc}) reads as
\begin{eqnarray}
\xi^3(1-\xi^4)\Omega(\xi)\Omega'(\xi)|^1 _0 = 0
\end{eqnarray}
We use a trial function as follows

\begin{eqnarray}
\Omega(\xi)=1-\beta\xi^2
 \end{eqnarray}
We expand (\ref{functional}) as a power series of  $O(\epsilon^2)$ (remember that $3<\alpha<4$), obtain
\begin{eqnarray}
\lambda^2(\beta,m^2,\alpha)=-\frac{20.4855 \left( m^2 \beta ^2-3 m^2 \beta +3m^2+6.8 \beta
   ^2-22.5 \beta +18\right)}{\left( \beta ^2-2.96617 \beta
   +2.41891\right) \epsilon ^2}\label{integral}
 \end{eqnarray}
Recall that near the critical point $T\rightarrow T_0$, the quantity
$\epsilon\propto <O_{\Delta}>$ is small \cite{kanno}. The minimum of
$\lambda^2(\beta,m^2,\alpha)$ exists at $\beta=0.304936$, with the
value
 $$|\lambda_{min}|^2=\frac{12.7445 \left(2.17818 m^2+11.7713\right)}{\epsilon ^2}$$
The minimum critical temperature $T^{Min}_c$ is
\begin{eqnarray}
T^{Min}_c=\frac{h_c}{\pi}=\frac{1}{\pi}\sqrt[3]{\frac{\rho\epsilon }{\sqrt{12.7445 \left(2.17818 m^2+11.7713\right)}}}\label{tc}
\end{eqnarray}
From (\ref{tc}) we observe that the mass of the scalar field must be $$m^2>m_c^2$$
Where $m_c^2=-5.40417$. It's in the range of the BF bound. This inequality gives a lower bound for the $m^2$.
 Further, it is not related to stability. It's just back to the real value of the $T_c$. It
's very interesting that we investigate the dynamical properties of
the superconductor near this lower bound. At first, we observed that
the critical temperature $T_c$ remains finite.

Specially for $m^2=-3$ we have
\begin{eqnarray}
T^{Min}_c=0.158047\sqrt[3]{\rho\epsilon }
\end{eqnarray}
The numerical value of the critical temperature \cite{plb2011} for $m^2=-3$, $\gamma=-0.06$
is $$T_c^{Min-Numerical}\approx 0.170\sqrt[3]{\rho}$$
 Our analytical result $$T_c^{Min-Analytical}\approx0.158047\sqrt[3]{\rho\epsilon }$$ is a
  lower bound for the numerical estimation. Since $\epsilon$ may be any value, if we put
   $\epsilon\approx1.25$, the results coincide to each other. In Figure.1 we plot the
$\frac{T_c}{\sqrt[3]{\rho }}$ as a function of $m^2$. From figure we
concluded that when the mass of the scalar field $m^2$ grows, the
value of the critical temperature increases.


\begin{figure}
\centering
 \includegraphics[width=6cm,angle=0] {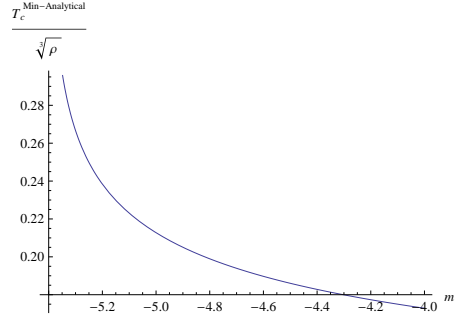}
  \caption{Plot of the $\frac{T_c^{Min-Analytical}}{\sqrt[3]{\rho }}$ as a function of the $m^2$ for $m_c^2< m^2\leq-4$. The lower and upper BF bounds are $\Big(m^{lower}_{BF}\Big)^2=-4,\ \ \Big(m^{upper}_{BF}\Big)^2=-3$.}
 \label{1.eps}
\end{figure}

We tabulated different values of the critical temperature in Table I. Specially it is
 interesting that when the mass of the scalar field reaches the BF bound i.e. $m^2=m_{BF}^2=-4$,
 the critical temperature  $T_c$ remains finite. The values of this table are in good agreement with
 the numerical data \cite{plb2011}.
  \begin{table}[ht]
  \caption{ Variation of the $\frac{T_c}{\sqrt[3]{\rho }}$ as a function of $m^2$ for  $ m_c^2< m^2\leq m_{BF}^2$.}
  \centering
  \begin{tabular}{|c| c |c| c |c|}
  \hline
  $m^2=-5.4$ & $m^2=-5$ & $m^2=-4.8$ & $m^2=-4.2$ & $m^2=m_{BF}^2$   \\ [0.5ex]
     \hline
      0.653067 & 0.142186 & 0.124354 &
0.0988134 & 0.0938797 \\
          \hline
      \end{tabular}
       \label{table:fluid}
       \end{table}

\subsection{linear relation between $<O_{\Delta}>$ and the chemical potential}

In this section we want to obtain the linear relation between
$<O_{\Delta}>$ and the chemical potential. For this reason, firstly
we rewrite the (\ref{eom22}) near the critical point $T\rightarrow
T_c$, with the solution (\ref{aproxpsi})
\begin{eqnarray}
\phi''+s(\xi)\phi'=\frac{( \epsilon\xi^{\Delta}\Omega(\xi))^2\Big(1+c_4  ( \epsilon\xi^{\Delta}\Omega(\xi))^2+c_{\alpha} ( \epsilon\xi^{\Delta}\Omega(\xi))^{\alpha-2}\Big)}{\xi^2(1-\xi^4)(1-24\gamma \xi^4)}\phi
\end{eqnarray}
Here $$s(\xi)=-\frac{(1+72\gamma \xi^4)}{\xi(1-24\gamma \xi^4)}$$
Near the critical point, we must keep only terms of order
$\epsilon^4$. Since $3<\alpha<4$, thus it's not necessary to keep
all the terms with coefficients. For example now we put $c_4\cong0$,
and only we keep the $c_{\alpha}$ term. We write the following
approximated solution \cite{cai}
\begin{eqnarray}
\phi(\xi)\approx\mu_c+\epsilon \chi(\xi)
\end{eqnarray}
Where $\chi(\xi)$ is a general function, with this auxiliary condition $\chi(0)=1$.
the equation for $\chi$ reads
\begin{eqnarray}
\chi''(\xi)+s(\xi)\chi'(\xi)\approx\mu_c\frac{\epsilon(\xi^{\Delta}\Omega(\xi))^2\Big(1+c_{\alpha} ( \epsilon\xi^{\Delta}\Omega(\xi))^{\alpha-2}\Big)}{\xi^2(1-\xi^4)(1-24\gamma \xi^4)}\label{chi}
\end{eqnarray}
The solution for $\chi(\xi)$ reads
\begin{eqnarray}
\chi'(\xi)=e^{-\int s(\xi)d\xi}(C+\epsilon \int j(\xi)e^{\int s(\xi)d\xi}d\xi)
\end{eqnarray}
Where
$$
j(\xi)=\mu_c\frac{(\xi^{\Delta}\Omega(\xi))^2\Big(1+c_{\alpha} ( \epsilon\xi^{\Delta}\Omega(\xi))^{\alpha-2}\Big)}{\xi^2(1-\xi^4)(1-24\gamma \xi^4)}\label{chi}\\, \  \
e^{\int s(\xi)d\xi}=\frac{-1+24\gamma \xi^4}{\xi}
$$
Now we have
\begin{eqnarray}
\chi'(\xi)=\frac{\xi}{24\gamma \xi^4-1}(C+\epsilon \int \frac{j(\xi)(24\gamma \xi^4-1)}{\xi}d\xi)
\end{eqnarray}
Finally we obtain
\begin{eqnarray}
\chi(0)=\frac{ \sqrt{6} \pi  C
   }{48  \sqrt{\gamma } }\label{chi0}
   \end{eqnarray}

 Also, we can expand the $\phi$ near $\xi=0$ as a series solution as the follows
\begin{eqnarray}
\phi\approx \mu-\frac{\rho}{h}\xi^2\approx \mu_c+\epsilon(\chi(0)+\chi'(0)\xi+\frac{1}{2}\chi''(0)\xi^2+...)
\end{eqnarray}
comparing the coefficients of order $\xi^0$ we get
\begin{eqnarray}
\mu-\mu_c\approx \epsilon\chi(0)
\end{eqnarray}
Using (\ref{chi0}) we obtain
$$
\mu-\mu_c\approx \epsilon\Big(
\frac{ \sqrt{6} \pi  C
   }{48  \sqrt{\gamma } }\Big)
   $$
Now we obtain
$$
\mu-\mu_c\approx <O_{\Delta}>\Big(
\frac{ \sqrt{6} \pi  C
   }{48  \sqrt{\gamma } }\Big)
   $$
Such linear relation between $<O_{\Delta}>$ and the chemical
potential has been discussed previously in s-wave and p-wave
superconductors \cite{cai}.

\section{Conclusions}
In this paper,we studied the analytical properties of the
Stuckelberg holographic superconductors with Weyl corrections, using
a variational method. Firstly we reduced the problem to a
variational Sturm-Liouville equation near the critical point. We
written a suitable functional, and using some trial functions, we
obtained the lower bound of the critical temperature $T_c$. We
showed that, when the expectation value of the dual operators with
conformal dimension $\Delta=3$ near the critical point $T=T_c$
remain small, we can calculate the $T_c$ easily. The expression for
$T_c$ is a function of the $m^2$. We obtained that for positive
$T_c$, we must have $ m^2>m_c^2$. This new lower bound on $m^2$  is
completely different from the same value of the lower bound obtained
from the stability. We discussed the relation between the $T_c$ and
$m^2$. We showed that when $m^2$ increases, the $T_c$ decreases and
the condensation becomes weaker. Further we show that near the BF
bound, i.e. when $m^2=m^2_{BF}$, the $T_c$ remains finite. It is
shown that there is no divergence near the BF bound for $T_c$.
Further we obtained that there is a linear relation between
$<O_{\Delta}>$ and the chemical potential.

\section{Acknowledgment}
The authors would like to thank \textit{Jian Pin Wu} (Beijing Normal
University-China) , \textit{Wen-Yu Wen} (Chung Yuan Christian
University-Taiwan) and   \textit{Mubasher Jamil} (Center for Advanced Mathematics and Physics -Pakistan) for reading manuscript and helpful discussions and
useful comments. We would like to thank anonymous referees for
giving useful comments to improve this paper.


\begin{thebibliography}{99}



\bibitem{maldacena} J. M. Maldacena, Adv. Theor. Math. Phys. 2, 231 (1998).
\bibitem{condencesd1}
C. P. Herzog, J. Phys. A 42 (2009) 343001 .
\bibitem{condencesd2}
 S. A. Hartnoll, Class. Quant. Grav. 26 (2009) 224002.
\bibitem{super1}
G. Policastro, D. T. Son,  A. O. Starinets, Phys. Rev. Lett.
87, 081601 (2001).
\bibitem{super2}
 P. Kovtun, D. T. Son,  A. O. Starinets, JHEP10 (2003) 064.
\bibitem{GR1}
S. A. Hartnoll, C. P. Herzog,  G. T. Horowitz, JHEP 12 (2008) 015.
\bibitem{GR2}
 G. T. Horowitz , M. M. Roberts, Phys. Rev.
D 78, 126008 (2008).
\bibitem{GB1}
R. Gregory, S. Kanno , J. Soda, JHEP10 ,010(2009)
.

\bibitem{GB2}
 Y. Brihaye , B. Hartmann, Phys. Rev. D 81, 126008 (2010).


 \bibitem{GB3}
R.-G. Cai, Z.-Y. Nie,. H.-Q. Zhang, Phys. Rev. D 82,
066007 (2010).
\bibitem{GB4}
 Q. Pan , B. Wang, Phys. Lett. B. 693 (2010)
159.

\bibitem{GB5}
 Q. Pan, J. Jing, B. Wang, JHEP 11 (2011) 088;
\bibitem{HL1}
R.-G. Cai , H.-Q. Zhang, Phys. Rev. D81, 066003(2010).
\bibitem{HL2}
 D. Momeni, M. R. Setare, N. Majd, JHEP 1105 ,118(2011),arXiv:1003.0376 [hep-th].
\bibitem{wen1}
E. Nakano, W.-Y. Wen, Phys.Rev.D78,046004(2008).

\bibitem{wen2}
D. Momeni, E. Nakano, M. R. Setare, W.-Y. Wen, arXiv:1108.4340v2 [hep-th].

\bibitem{epl1}
M. R. Setare, D. Momeni, EPL, 96 (2011) 60006,arXiv:1106.1025.

\bibitem{born1}
Y. Liu, Y. Peng, B. Wang,  arXiv:1202.3586 [hep-th].

\bibitem{born2}
 Q. Pan, J. Jing, B. Wang,  Phys. Rev. D 84, 126020 (2011).
\bibitem{cs}
 G. Tallarita, S. Thomas, JHEP 1012:090,2010.

 \bibitem{maxwell}

Q. Pan, J. Jing , B. Wang  Phys.Rev. D84, 126020 (2011) .

\bibitem{cai2012}
 Y.-Q. Wang, Y.-X. Liu, R.-G. Cai, S. Takeuchi, H.-Q. Zhang, arXiv:1205.4406 [hep-th].
\bibitem{pwave1}
S. S. Gubser, S. S. Pufu, JHEP 11 (2008) 033.
\bibitem{pwave2}
 C. P.
S. Herzog,  S. Pufu, JHEP 0904 (2009) 126.
\bibitem{pwave3}

 M. Ammon ,
J. Erdmenger ,
V. Grass, ,
P. Kerner ,
A. O'Bannon
, Phys. Lett. B 686 ,192(2010) .


\bibitem{herzog}
C.~P.~Herzog,  Phys.\ Rev.\ { D81}, 126009 (2010).


\bibitem{kanno}
S. Kanno, Class. Quant. Grav. 28, 127001(2011) .

\bibitem{analytic1}
H.-B. Zeng, X. Gao, Y. Jiang, H.-S. Zong, JHEP 05 (2011) 2.
\bibitem{analytic2}
Q. Pan, J. Jing , B. Wang , S. Chen ,JHEP, 1206, 087(2012).


\bibitem{weyl}
J.-P. Wu, Y. Cao, X.-M. Kuang, W.-J. Li, Phys. Lett.
B697, 153 (2011).
\bibitem{mpla}
D. Momeni,  M.R. Setare, Mod. Phys. Lett. A,  26, 38,2889 (2011), arXiv:1106.0431 .
\bibitem{jian2010} J.-P. Wu, arXiv:1006.0456 [hep-th].
\bibitem{epl2} D. Momeni,N. Majd, R. Myrzakulov, EPL, 97 ,61001(2012),arXiv:1204.1246 [hep-th].
\bibitem{plb2011}
D.-Z. Ma, Y. Cao, J.-P. Wu ,Phys. Lett. B 704, 604(2011) .


 \bibitem{qc}
 I. T. Drummond , S. J. Hathrell, Phys. Rev. D 22, 343 (1980).

\bibitem{BF}
P. Breitenlohner , D. Z. Freedman,  Phys. Lett.B. 115,197 (1982) .
\bibitem{cai}
R.-G. Cai, H.F. Li, H.Q. Zhang, Phys.Rev.D83:126007(2011).

\end{thebibliography}
\end{document}